\documentclass{fmj2010}

\def\fermi{\textit{Fermi}}

\begin{document}
   \title{Probing the Physics of Gamma-Ray Blazars with Single-Dish Monitoring Data}

   \author {M. F. Aller, P. A. Hughes \& H. D. Aller}
   \institute{Department of Astronomy, University of Michigan, Ann Arbor, MI, 48109-1042,
 USA}

   \abstract{
In the 1990s a comparison of sparse EGRET measurements with single-dish
flux density monitoring from the Mets\"ahovi and UMRAO programs established
a temporal connection between the onset of flaring at radio band and the occurrence
of gamma-ray activity. Correlations between the emergence of new VLBI components
from the core, flares in linearly polarized radio flux, and gamma-ray activity in bright 
EGRET-detected blazars supported a picture in which the gamma-ray and the radio band
emission arises in the same shocked region of the jet, with the high energy emission produced
via inverse Compton scattering by the synchrotron-emitting electrons in the jet. Quantitative tests
of this scenario, however, were hampered by insufficient temporal sampling of the data and the 
simple nature of the models adopted. The extensive data from \fermi\ coupled with the wealth of
well-sampled radio band data from old as well as new programs such as the F-GAMMA project
now permit statistical studies for large numbers of sources, including weak HBLs, and detailed
analyses of individual highly-active class members. I summarize progress in understanding the
origin of the gamma-ray emission using these new measurements. I focus on three areas: attempts 
to isolate the physical site of the high energy emission using time delay information; investigation
of the emission process using the characteristics of the variability; and quantitative tests of the
shock model picture using high-time-sampled multifrequency linear polarization data, VLBP
imaging, and new models of propagating oblique relativistic shocks incorporating detailed
radiative transfer calculations.
   }

\titlerunning{Single Dish Monitoring}
\maketitle 

\section{Introduction}

In spite of the limitation provided by the relative low sensitivity and poor time sampling (detection statistics only) from EGRET,
these measurements set the framework for our current understanding of the MeV-to-GeV $\gamma$-ray emission from AGN.
The enormous increase in sensitivity and the use of large-area sky scan mode rather than pointings made possible by the
$>$ 2 sr field of view accessible to the Large Area Telescope (LAT), the principle instrument for studying blazars onboard 
 \fermi\, has resulting in $\gamma$-ray light curves with sufficient time sampling to track the variations on
intraday time scales during extreme strong flares, and on daily-to-weekly time scales for
tens of bright AGN in the energy band 20 MeV to $>$ 300 GeV. To match these data, a new generation of
radio band monitoring programs
has emerged, providing well-sampled light curves for hundreds of sources. These enable
investigators to carry out both statistical studies of large samples of AGN to low radio-band flux limit, and 
detailed studies of individual objects using multifrequency data provided in support
of the \fermi\ mission. Using well-established procedures, including cross-correlations
of intra-band data permitting direct comparison of the activity across bands and investigations of
 the character of the variability itself through analyses which probe the
emission process, detailed studies of the variability properties
in {\it both} bands can be carried out. These comparison rely on
single dish monitoring data which provide temporal resolution complementing the 
spatial resolution provided by monthly-to-less-frequent VLBI imaging measurements. I summarize the
current status of these monitoring programs and review what they can tell us about the site of and mechanism for the
generation of $\gamma$-ray emission from AGN detected by {\it Fermi}.

\section{The EGRET era: setting the stage}

The launch of the Compton Gamma Ray Observatory (CGRO) in April 1991 provided us with a new view of familiar blazars.
One of the early discoveries from EGRET (the instrument onboard operating above 30 MeV) was the identification of a population
 of GeV-$\gamma$-ray-emitting blazars,
positionally associated with well-known radio-bright AGN. Comparison of activity monitored 
in the millimeter band  (37 \& 22 GHz) in a sample of $\approx$ 70 bright sources combined with the EGRET detections
 in the first EGRET catalogue (Fichtel et al. 1994), the results of an all sky survey
carried out from 1991 April to November 1992,
 established a  statistical connection between enhanced radio band activity 
and  detection in the energy band E$>$100 MeV (e.g. Valtaoja \& Ter\"asranta \cite{valtaoja95}).
 Note that only a handful of photons were detected by EGRET 
above 10 GeV (Thompson \cite{thompson06}), and that much information was extracted based on only a small number of
photons in these early studies.
Even with the paucity of EGRET data and the relatively small number of sources
detected, it became quickly apparent that these $\gamma$-ray-detected
sources were predominantly radio-bright, compact, core-dominated
blazars with high brightness temperatures (Valtaoja \& Ter\"asranta \cite{valtaoja96}). Further, 
a temporal association between the occurrence of an EGRET detection and the rise portion of the radio 
band flare (start of flare to maximum) was established for
individual events using both the  millimeter (Valtaoja \& Ter\"asranta \cite{valtaoja96})
and the centimeter band (Aller, Aller \& Hughes \cite{aller96}) total flux density
monitoring data.  An example plot illustrating this result is shown in Figure 1 which includes EGRET phase 1+2+3
$\gamma$-ray data and  Mets\"ahovi monitoring measurements obtained during this time
period (see also L\"ahteenm\"aki \&
Valtaoja \cite{lahteenmaki99}). Because of the severe undersampling of the EGRET data, however, it was not possible to carry
out cross-correlation analyses between the two bands. 

\begin{figure}
\centering
\vspace{230pt}
\includegraphics{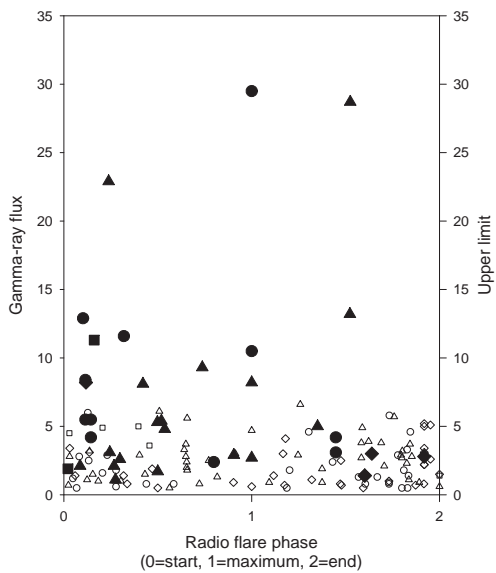}
\caption{Time of occurrence of $\gamma$-ray detection as a function of flare phase based on Mets\"ahovi data. Figure courtesy A.
 L\"ahteenm\"aki}
        
\end{figure}

Attempts to correlate $\gamma$-ray photon flux and radio flare peak flux or radio luminosity
yielded mixed results that ranged from no clear correlation (Aller, Aller, \& Hughes \cite{aller96})
to claimed correlations. This discrepancy stemmed from a variety of causes including the use of non simultaneous data,
dependence of the luminosity on redshift, and time delays between the radio and $\gamma$-ray activity produced by
opacity in the radio jet (M\"ucke et al. \cite{mucke96}). An example flux-flux correlation
is shown in Figure 2 based on detections from EGRET and monitoring observations from the Mets\"ahovi program 
(see L\"ahteenm\"aki et al. \cite{lahteenmaki97}). 

Studies of a handful of individual, exceptionally-bright sources for which the binned 
EGRET measurements yielded multiple detections (time-averaged photon flux values as a function of time),
in combination with radio band light curves
and VLBI imaging data, confirmed the physical association between flaring in the two bands identified statistically.
Analysis of the data for sources such as NRAO 530, confirmed the
general trends but found cases of $\gamma$-ray detections during declining or constant
millimeter-to-entimeter band flux (Bower et al. \cite{bower97}) which deviated from the
general trends.
 
The link to flaring, and subsequently the identification of a specific flare 
phase established a connection to the development of individual radio band outbursts. Since the radio band variations are generally
attributed to shocks in the radio jet, this connection led investigators to the conclusion that shocks might also play a role in
the generation of the $\gamma$-ray flaring (e.g. Valtaoja \& Ter\"asranta \cite{valtaoja96}).

\begin{figure}
\centering
\vspace{230pt}
\includegraphics{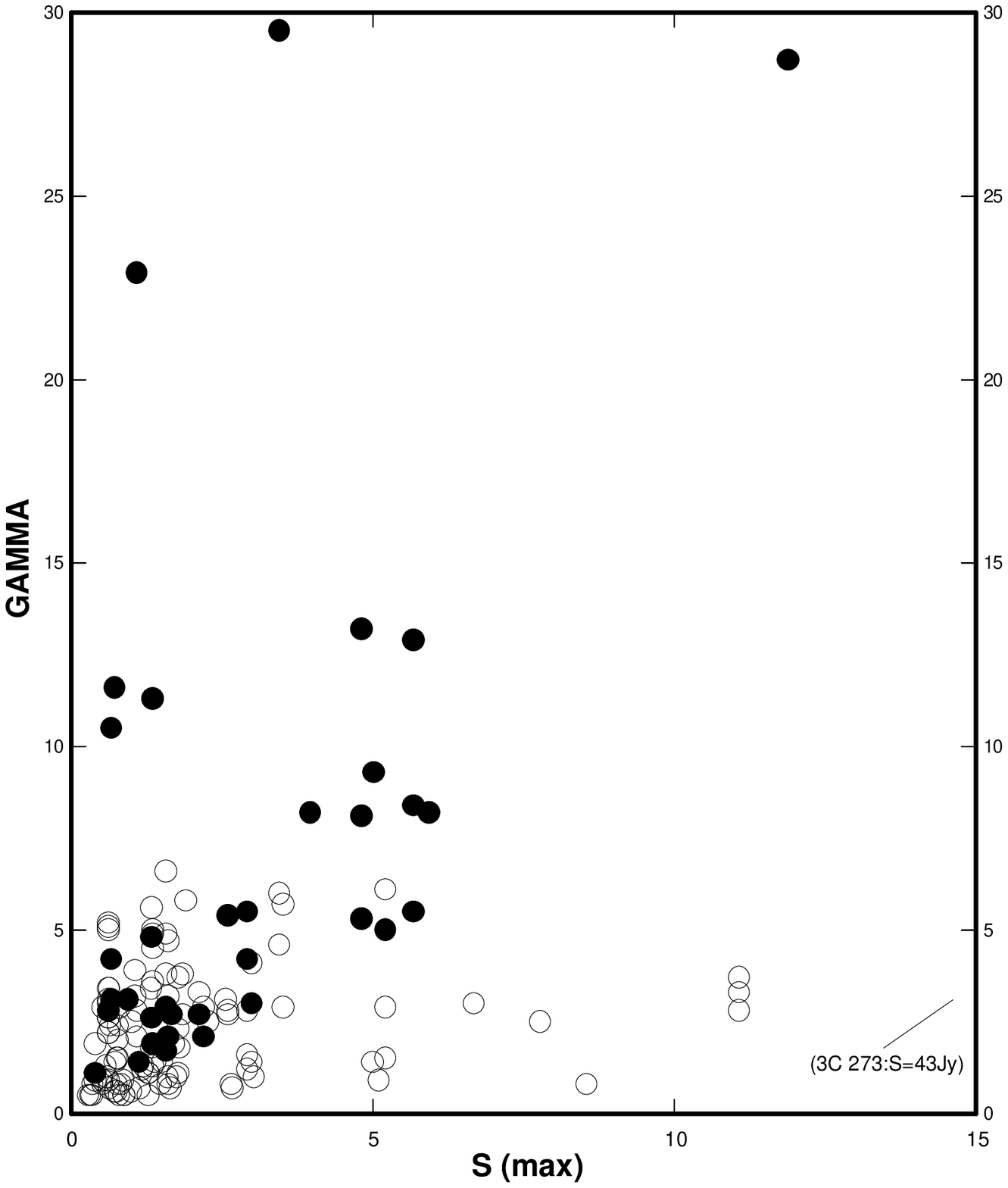}
\caption{$\gamma$-ray flux versus Mets\"ahovi peak flare flux. Figure courtesy A. L\"ahteenm\"aki.}
        
\end{figure}

Linear polarization monitoring data track temporal changes in magnetic field structure. In well-resolved events, such
data can be used to search for the signature of a shock: a swing in electric vector position angle (hereafter EVPA) and an
increase in the fractional linear polarization (hereafter LP). These changes are associated with the compression of the 
initially tangled magnetic field  during the passage of the shock.
The 22 \& 43 GHz VLBA imaging data for 42 $\gamma$-ray-bright blazars obtained from the Jorstad \& Marscher program,
in combination with UMRAO linear polarization monitoring data, were used to identify a correlation between
high $\gamma$-ray state and an increase in polarized radio band flux
supporting a picture in which the radio emission and the $\gamma$-rays originate in
the same shocked region of the jet, with the $\gamma$-ray emission produced by inverse Compton scattering 
in the thin forward region of the shock where the electrons are accelerated (Jorstad \cite{jorstad01}).
This scenario placed the $\gamma$-ray emission site parsecs far from the central engine
and downstream of the radio band core.

The majority of the EGRET-detected sources were highly luminous HPQs. With only a few exceptions,
these were the very same radio-bright sources which had been included for decades in the Mets\"ahovi
and UMRAO flux monitoring programs. We now know that this was a selection effect resulting from the limited
sensitivity of EGRET.

\section{Single dish radio band monitoring during the {\it Fermi} era}

In spite of the progress made in understanding the origin of the $\gamma$-ray emission using the EGRET data,
answers to many questions remain in areas where single dish monitoring data can provide insights.  Specific
questions which can be addressed using these data are:
\newline 1. Where within the jet is the $\gamma$-ray flare produced? This question can be examined using
 cross identification of flares in the high energy and radio bands. A combination of single dish monitoring
and VLBI imaging can be used to localize the physical site of the high energy emission using the radio core
as a fiducial reference point.
\newline 2. What emission mechanism is responsible for the generation of the $\gamma$-ray emission? This can
 be explored using SEDs based on contemporaneous broadband measures, including single-dish total flux density
data at multiple epochs. These trace the evolution in the $\nu F_{\nu}$ versus $\nu$ plane.
\newline 3. What is the mechanism for  acceleration of the particles? Tests for the presence of
 shocks during $\gamma$-ray flaring using a combination of data and modeling can be carried out.
\newline 4. What special conditions are present in the jet during broadband flaring? Information on
jet properties during individual flares and on changes in jet conditions can be derived using single dish measurements.

\fermi\ scans the sky every 3 hours providing data which can be analyzed to study the variability in the MeV-GeV band on
near-daily to weekly time scales.
In support of this program several new programs as well as continuing old ones
 monitor the total flux density of AGN.  The
salient features of several of these programs providing data in the centimeter-to-millimeter
band  for sources
north of approximately -30$^{\circ}$ are summarized in Table 1. The first
three entries form the F-GAMMA alliance which works in tight coordination. The OVRO program 
provides time-sampling at a single frequency, 15 GHz, temporally matched to that attained using {\it Fermi}. 
Observations in this band are well-suited for tracking variability since the amplitude of the
variability is  relatively large compared to that at lower frequencies, and the measurements are
relatively free from tropospheric effects which increase with higher
frequency. The UMRAO and Mets\"ahovi programs provide long term histories for smaller
numbers of sources and with less-frequent cadence.  The AGN population now detected by \fermi\ includes a 
substantial population of radio-weak sources belonging to the HBL class. These radio-weak sources
can only be monitored using single dish instruments with large collecting areas such as
the OVRO 40-m telescope and the Effelsberg 100-m instrument.  Spectral data is provided by Effelsberg/IRAM
and the RATAN-600 programs. Together the data base of observations includes both temporal and spectral coverage
for a large number of sources. Multifrequency data, including linear polarization measurements, are themselves
important for identifying both self-absorption and Faraday effects. 

\begin{table}
\caption{Current Single Dish Monitoring Programs}
\label{ Programs:1}
\centering
\begin{tabular}{l c c c }
\hline\hline
Prog.      & freq. (GHz) & sampling   & size/advantage \\
\hline
\noalign{\smallskip}
OVRO & 15   & 2-3/week  & $>$1000                \\
     &      &           &  many sources; low S \\
\hline
Effsbg.  & 2.64--43  &  monthly & $\approx$60    \\
         &              &     & spectra           \\
\hline
IRAM     & 86--270  &  monthly & $\approx$60    \\
         &          &          & inner jet       \\
\hline
UMRAO & 4.8, 8, 14.5  & 1-2/week & 35 in core grp. \\
        &             &          & mf; includes LP   \\
\hline
Mets\"ahovi & 37      & monthly   & $\approx$100    \\
            &        &            & inner jet \\
\hline
RATAN-600   &  1-22  & 2-4/year    & 600           \\
            &        &     & spectra    \\
\hline

\end{tabular}

\end{table}

\section{Is the broad band activity causally related? Radio band-$\gamma$-ray correlations}

Evidence in support of a statistically-significant correlation between the fluxes in the two bands
would support the hypothesis that the emissions are related. Results based on observations from the OVRO program
for 49 sources in the \fermi\ LAT three month catalogue are shown in Figure 3 from the work of Richards et al.
(\cite{richards09}). The data shown are averages over the three month period August 4 -- October 30, 2008. 
A correlation is apparent visually. A formal analysis gives a
correlation coefficient of r=0.56; Monte Carlo simulations indicate  a 
chance probability of this correlation of only 5$\times10^{-4}$.

\begin{figure}
\centering
\vspace{230pt}
\includegraphics{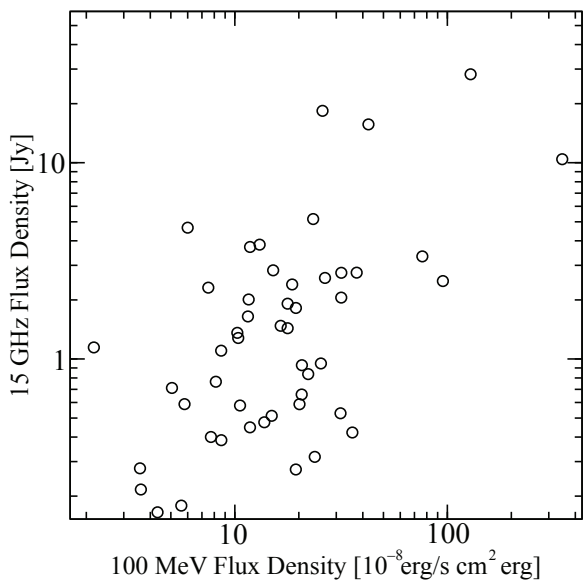}
\caption{\fermi\ LAT flux density in the 100 MeV energy band versus OVRO 15 GHz flux density based on data obtained
 in the first 3 months of operation (Richards et al. \cite{richards09}).}
\end{figure}

Figure 4 shows the flux-flux correlation obtained using results from MOJAVE for sources in the 
high-confidence 3-month LAT list (Abdo et al. \cite{abdo09}) with 
radio flux density $>$0.2 Jy and $\delta\geq$-30 degrees. The 15 GHz fluxes shown
include both MOJAVE data (total flux density over the entire VLBA image) and supplementary UMRAO and RATAN-600 fluxes (shown by
open circles). This work found a correlation
between the $\gamma$-ray flux in the first three months of \fermi\ operation and these
quasi simultaneous 15 GHz fluxes, and identified that the radio jet was in an enhanced state near the time
of strong $\gamma$-ray emission for sources in this group (Kovalev et al. \cite{kovalev09}).

\begin{figure}
\centering
\vspace{230pt}
\includegraphics{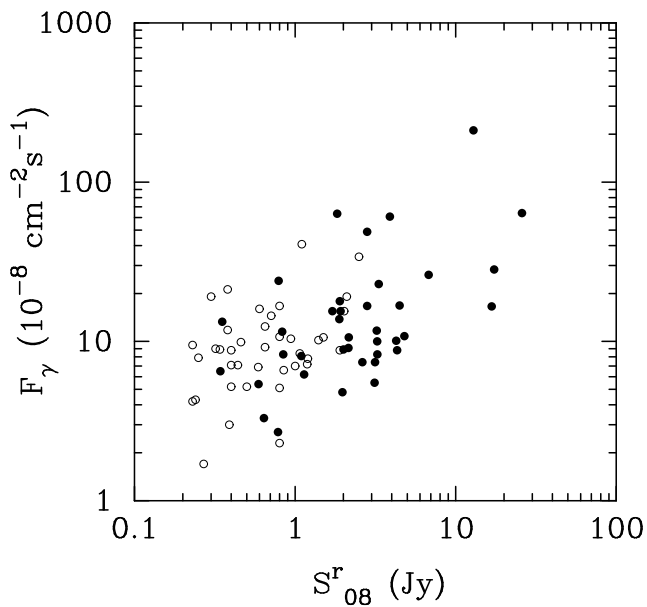}
\caption{Average $\gamma$-ray photon flux in the 100 MeV-1 GeV energy band  versus 
15 GHz flux for sources in the MOJAVE sample.}
        
\end{figure}

\section{Characterization of the Variability}

Variability time scales, variability amplitude, quasi-periodicities,  and indicators
of the noise process producing the variability can be extracted from the
detailed study of light curves. Common analysis tools include
structure functions, PDSs, and wavelets.
Useful measures of the degree of variability include the fluctuation index and the normalized
excess variance.  Similar characteristics in the variability of the two bands would support a related origin.
However, a sufficiently long data train is required to capture the characteristics of the variability. 

\subsection {Results}
Structure functions, autocorrelations and power spectra are related measures of
the distribution of power versus time. Structure functions can provide both a characteristic time scale
and a ``measure'' of the noise process; the former comes from the time lag where the curve plateaus and the
latter from the slope of the curve. 
 First order structure functions were computed for the best observed (i.e. most variable) 
sources in the UMRAO
program (Hughes, Aller, \& Aller \cite{hughes92}); the turnover yielded 
a characteristic variability time scale of 2 years while the slope, generally near 1, was consistent
with shot noise. Because of the length of the data train in the radio band, several events in each source
were captured in general; in a few sources, e.g. 3C~84 and 3C~454.3, the variability
was dominated by a single event, and in these cases the slope of the structure function was 1.55 -1.6.

In the case of the $\gamma$-ray data, nearly two years of data have
accumulated. A study containing light curves of 106 sources showing weekly averages of the data
(E$>$300MeV) obtained over the first 11 months of operation has recently
been submitted for publication which includes analysis using structure functions and PDSs.
(Abdo et al. \cite{abdo10c}). Characteristic structure function power indices are in the
range 1.0 to 1.6. Higher values are found for 0235+164 and 3C~454.3; in these sources the light curves
are dominated by a single large event. The light curves of these two objects in the radio band show
the same characteristic behavior during this time period.

\subsection{Time-Delay Studies}
The site of the $\gamma$-ray emission can in principle be identified
relative to the emission site for the radio emission from time delays by using
well-sampled light curves to look for correlated activity.
If similar features can
be unambiguously identified, the associated time delays, provide
information on the location of the $\gamma$-ray emission site. In the radio band this can
be referenced to the radio core at 43 GHz identified from VLBI imaging.
Monitoring observations from the OVRO program have traced the total flux density variability at 15 GHz
in hundreds of sources with a time sampling well-matched to that provided by the LAT. Cross-correlation of the OVRO
15 GHz and early \fermi\ data for a few representative
cases are presented in Max-Moerbeck \cite{max09}. These few examples illustrate that a range of
behavior patterns are present in the time delays. While activity is clearly associated across the spectrum,
unambiguously identifying individual events across the spectral bands is extremely
difficult except during rare, large-amplitude, resolved flares. In other cases, frequency-dependent factors
including self-absorption and opacity, differ from source to source and
affect the time delay results.  An underlying problem in such comparisons is that  
persistent trends may not exist from source to source, or even in the same source from epoch to epoch;
this lack of persistence has recently been quantitatively examined in the optical to $\gamma$-ray regime
and attributed to variations in a number of parameters which are important in the emission process
(B\"ottcher \& Dermer \cite{boettcher10}).


\subsection{Periodicity}
  Quasi-periodicities have been identified in several sources observed in the UMRAO program using a variety of
analysis tools. The strongest evidence for periodicity in the centimeter band for any source is for OJ~287.
Visual inspection of the light curve for this source shown in Figure 5 reveals a long term trend with
 superimposed flares. Quasi periodic
behavior has been identified in this source using wavelets (Hughes, Aller, \& Aller \cite{hughes98}). The
total flux density amplitude is currently approaching the maximum value attained over the UMRAO multi-decade time period,
and the source is also bright at $\gamma$-ray band (see Figure 8). No evidence for periodicity has been found
(yet) using the \fermi\ light curves over the first 11 months of operation, and it will be very interesting to
see what analyses of longer data trains reveal.

\begin{figure}
\centering
\vspace{230pt}
\includegraphics{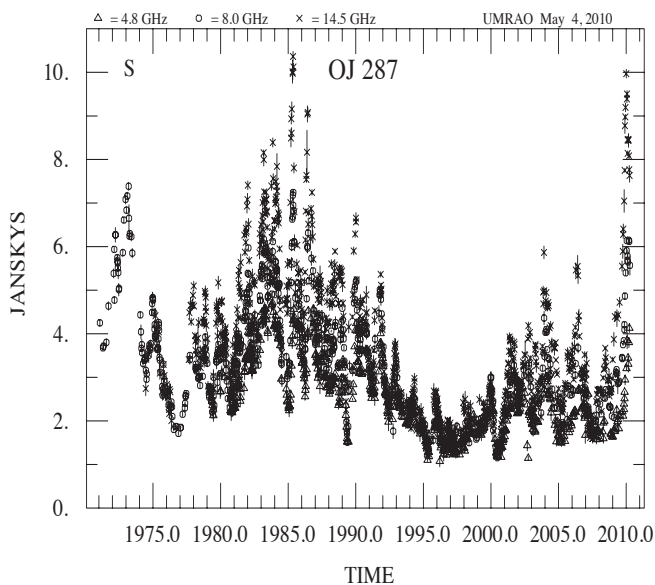}
\caption{Weekly averages of the total flux density for OJ 287 from the UMRAO program showing
 triple-frequency data obtained over 4 decades.
}
\end{figure}

\section{Emission Process: studies of SEDs using single dish data}
With the high time sampling now available, the evolution of the SED during individual events can be followed; such data
can provide insight into the emission mechanism using modeling. The low frequency hump is generally attributed
to synchrotron self-Compton emission produced in the jet, while the second  (high frequency) hump is often
attributed to  a complex mix of components (e.g. Abdo et al. \cite{abdo10d}). 
An example SED during flaring is shown for 3C~279 (Abdo et al. \cite{abdo10a}) in Figure 6.
In the radio band the SEDs at the two epochs overlay, while significant differences in the two time
intervals are apparent in all other bands.

\begin{figure}
\centering
\vspace{230pt}
\includegraphics{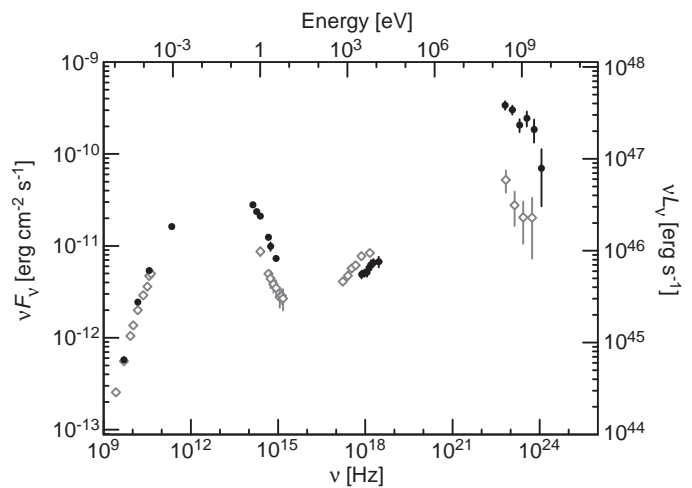}
\vspace{-25pt}

\caption{Evolution of the SED during $\gamma$-ray flaring. The filled symbols denote data obtained during the
first five days of a sharp $\gamma$-ray flare.  The unfilled triangles show data obtained about 70 days later
during a time when the $\gamma$-ray flux was relatively quiescent. In the radio band, the total flux density showed
a small systematic decline. The figure was kindly provided by M. Hayashida.
}
        
\end{figure}

\begin{figure}
\centering
\vspace{230pt}
\includegraphics{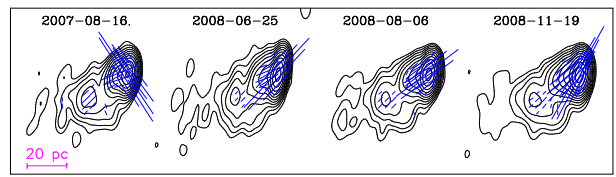}
\vspace{-150pt}

\caption{EVPA swing detected by MOJAVE in 1502+106 at epochs bracketing the 2008 August flare. The
most significant changes occurred during the time interval between the first and the 
second images shown. Figure provided by Y. Y. Kovalev.
}
        
\end{figure}

\section{Mechanism for particle acceleration: Shocks in the radio jet during $\gamma$-ray flaring}

Independent evidence in support of a scenario in which shocks propagating outward in the jet flows are the origin of the observed
blazar activity in the radio through optical spectral regions has come from both model fits of broadband spectral data (e.g.
Marscher \& Gear \cite{marscher85}) and of fits to centimeter band multifrequency flux and linear polarization data
during outburst using transverse
shock models (e.g. Hughes, Aller, \& Aller \cite{hughes89}). One possibility for the origin of these 
disturbances is Kelvin-Helmholtz-induced
instabilities. Both detailed hydrodynamical simulations of jet flows and stability analyses have confirmed
that such instabilities develop naturally and commonly in the kinetically-dominated portion of
the jet; current driven instabilities have also been considered and may
dominate in the magnetically dominated regime near the central engine (e.g. Hardee \cite{hardee06}). 
Shocks have been cited as a means of particle acceleration resulting in flaring in 
the $\gamma$-ray band, but searches for specific
evidence of shocks during $\gamma$-ray flaring have not been widely conducted.  In principle, shocks
in the parsec scale regime of the jet can be identified by searches for the expected swing in the orientation of the EVPA, 
apparent in consecutive VLBP images, but in general the VLBI sampling is too sparse to `detect' these
transitions. An example of this kind of flip in EVPA  is shown in Figure 7 which displays the MOJAVE 15 GHz images
bracketing flaring detected by \fermi\ `near' to
a major flare which occurred in 2008 August (Abdo et al. \cite{abdo10b}). In general the VLBA imaging data is too sparsely sampled 
 to permit following a shock event in detail. Single dish monitoring of the linear polarization
can provide the temporal resolution, but the shock signature will be smoothed out if there are 
contributions to the emission from many competing, evolving components. In sources dominated
by a single or only a few components, the signature of a shock appears in the single dish monitoring data 
as as ordered swing in EVPA (through 90 degrees in the special case of a transverse shock) combined with an increase in the
degree of linear polarization. In the general case, shocks are expected to be oriented obliquely to the flow direction.
The emission properties of such shocks have not been explored in detail in past studies.

\begin{figure}
\centering
\vspace{230pt}
\includegraphics{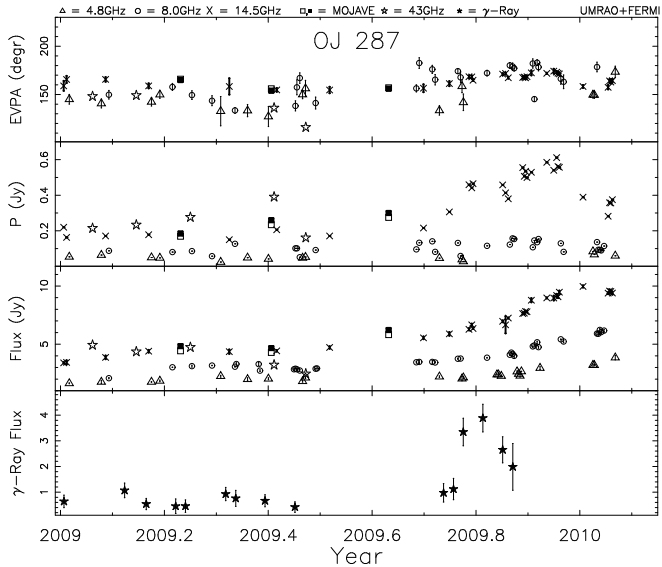}
\caption{Shock signature during flaring in OJ~287 in late 2009. From bottom to top weekly-averaged $\gamma$-ray photon flux
 and daily averages of radio band total flux density
and LP (fractional linear polarization and EVPA), For comparison
 VLBA fluxes at 43 GHz (denoted by stars) and a 15 GHz from MOJAVE (source-integrated values
filled squares; core only unfilled). $\gamma$-ray fluxes (units photons/sec/cm$^2\times$10$^{-7}$) were kindly provided
by S. Jorstad (private communication).}
\end{figure}

\begin{figure}
\centering
\vspace{230pt}
\includegraphics{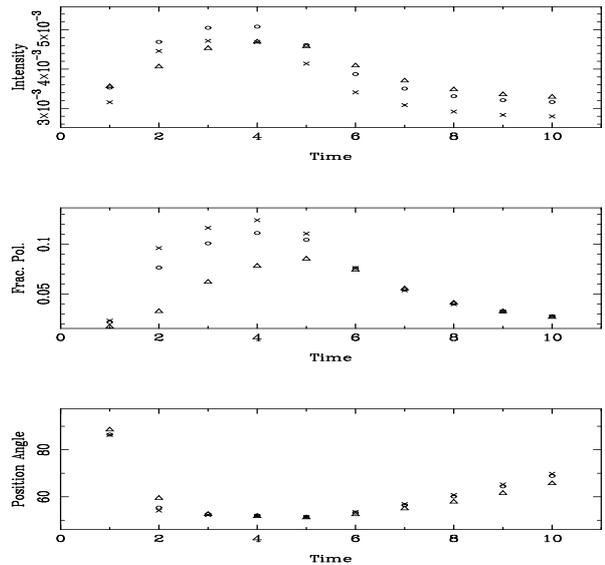}
\caption{Simulated light curves from radiative transfer calculations showing from top to bottom
total intensity, fractional polarization, and EVPA for three frequencies.
The Lorentz factor of the flow=2.5, the Lorentz factor of the shock= 6.7,  and the
viewing angle=10 degrees. Other model parameters are given in the text.}
\end{figure}

As part of the Michigan monitoring program, we are currently obtaining single dish monitoring observations
with a cadence of 1-2 times weekly of both the
total flux density and the linear polarization for a small sample of sources known to be flaring
in the $\gamma$-ray band and sufficiently bright (S$\geq$0.4 Jy) in the radio band to look for a shock signature
during times of $\gamma$-ray flaring. The program sources (3C 66A, 0235+164, 3C~84, 0420-014, 0454-234,
0528+134, 0716+714, 0727-115, 0805-077, OJ~287, 3C~273, 3C~279, 1308+326, 1502+106, 1510-089, 1633+382, OT~081,
2022-077, BL~Lac, CTA~102, \& 3C~454.3) are all members of MOJAVE, and 15 of the 21 are in the
Boston U. 43 GHz blazar program led by A. Marscher and S. Jorstad.  These VLBI data provide crucial
complementary information on flow direction relative to the magnetic field
orientation and a pattern speed characterizing the jet flow, as well as a means for identifying
sources with relatively simple structure. Example light curves obtained
for one of our program sources, OJ~287, are shown in Figure 8. The LP exhibits the expected
oblique shock signature -- a swing in EVPA through about 40$^{\circ}$  and an increase in fractional linear polarization from
near 0 to a maximum value approaching 8\% -- temporally associated with the $\gamma$-ray flaring in
late 2009. The LP
swings in EVPA that we have found in this and other sources take place over a time period of order one to two months
and require frequent measurements to properly track the variability. In addition to the UMRAO
light curves, we show in this figure MOJAVE 15 GHz  and BU 43 GHz measurements for comparison. 
These illustrate the excellent agreement between the single dish and VLBA results and graphically show the sampling
rate relative to the rapidity of the event itself. Note that in spite of the self-absorption apparent in the UMRAO
light curve in the 4.8-14.5 GHz band, that the spectrum is relatively transparent in the 15-43 GHz domain
as indicated by the nearly equal flux amplitudes in total flux and linear polarization. 

To explore the effect of the various input parameters on the predicted light curves and eventually to use
the models to analyze specific events, we have developed new radiative transfer models which incorporate
oblique shocks. These models assume an extreme relativistic equation of
state and are determined principally by two free parameters: the shock compression and the direction of
the shock (forward or reverse). The direction is important for time delay considerations.
From modeling of such events, information can be extracted about the conditions in the jet during $\gamma$-ray flaring;
changes in these conditions from event to event can  be identified and explored as more data accumulates. 

We show in Figure 9 representative, simulated light curves from this work. In the example shown, a shock oriented at
an obliquity of 45 degrees to the flow direction has been 
introduced into the relativistic flow at t=0. A forward-moving shock and a compression of 0.7 have been assumed.
The model predicts the general features apparent in the data (see Figure 8): a total flux outburst, an increase in the linear
polarization to near 10\%,  a swing in EVPA through about 40 degrees, and the general spectral behavior.
Figure 10 contains a series of images showing the evolution of the simulated structure. Images
such as these will be generated for comparison with VLBP imaging data to serve as model constraints
in the fitting procedure.

\section{Future Work}
Comparison of the variability detected in the \fermi\ and radio spectral bands continues to support a
generic scenario in which the same disturbance is responsible for both the radio band flares
and the $\gamma$-ray activity, most likely a shock, located
near the radio core in the spine of the radio jet (e.g.
A. Pushkarev, This Workshop). However, some
nagging exceptions to this simple picture remain. For example it does not account for the
presence of rapid (hourly) variability in the $\gamma$-ray emission detected during strong flares
 by \fermi\ (Tavecchio et al. \cite{tavecchio10}),
and known from earlier EGRET results. This rapid variability implies a small source size and
can most easily be explained by an emission site near the central engine.
Such results call into question the completeness of a picture which attempts to explain all emission
as originating at one site and by one mechanism. Exceptions to the rule need to be reconciled with the
generally-adopted picture.

The specific conditions in the jet leading to $\gamma$-ray flaring have not been isolated. 
These can be examined using a combination of the unique temporal resolution provided by single
dish monitoring, spatial information from VLBI studies, and detailed modeling of the emission
patterns. Model development, such as the shock modeling described here, is only in its very early stages.
Refinement will include the development of more sophisticated models and fits of these models to specific flares.

Data from large telescopes such as the OVRO 40-m and the Effelsberg 100-m dishes probe the variability properties of pure BL
Lacs and the large population of radio-weak HBLs, with sub-weekly cadences which can not be
provided by VLBI imaging alone. Detailed analyses of these sources may reveal other properties and behavior patterns
which need to be included in the general picture. Recent work from Mets\"ahovi suggests that even the RBLs may have
different emission properties characterized by $\gamma$-ray photon fluxes which are weak relative to those from other source
types and uncorrelated with radio-band flaring (Leon-Tavares, This Workshop).

\begin{figure}
\centering
\vspace{230pt}
\includegraphics{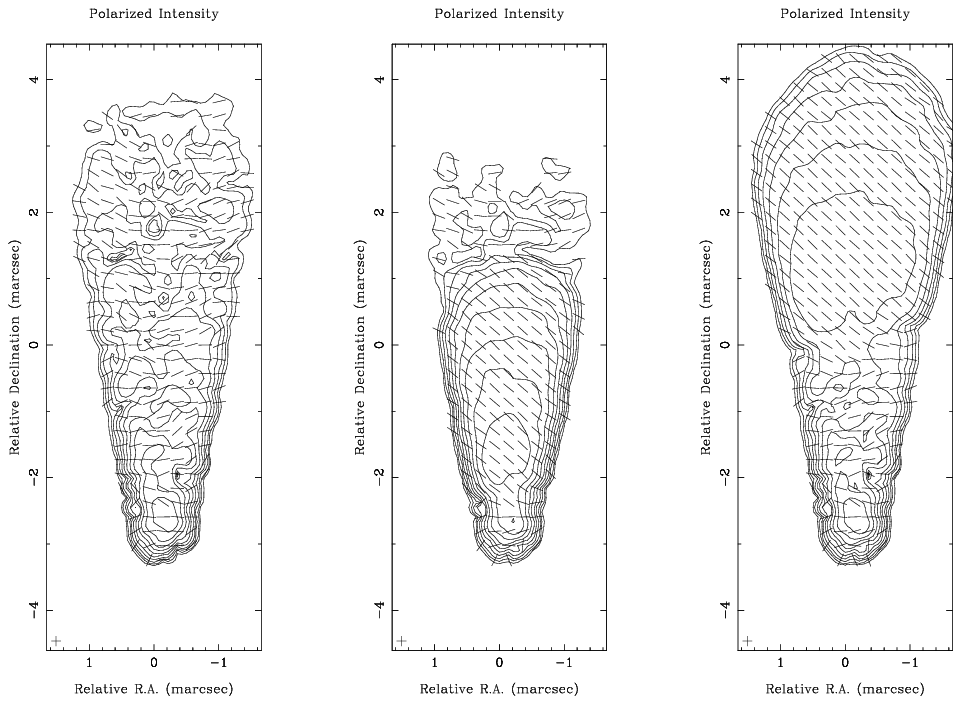}
\vspace{-60pt}
\caption{A sequence of simulated images from radiative transfer calculations showing polarization contours and
EVPA orientation as a function of time.}
\end{figure}

Long-term light curves in both the optical and radio spectral bands, have identified that 
there are changes in the character of the variability on time scales of several years to
decades. Searches for associated changes in the character of the radio band and $\gamma$-ray activity as the
\fermi\ data accumulates would be further support of an intimate connection between the emission in
these two regimes. Such studies will rely on historical data from long-term programs which
provide consistently acquired data at a common observing frequency in combination with the data
acquired during the operation of {\it Fermi}.

The first indications of an association between the emission in the radio and $\gamma$-ray bands
came from observations of a few extreme sources and limited photon statistics by EGRET. This
general picture can be refined and filled in using the wealth of data now available from
both the single dish data in the radio band and \fermi\ data. This will not an easy task because 
of the complexity of the emission patterns and the dependence on many parameters, but exciting, new
results will surely result enhancing our understanding of Blazars during the \fermi\ era.

\begin{acknowledgements}

This work was supported by NASA fermi cycle 2 grant NNX09AU16G, The long term operation of UMRAO has been
made possible by funds from the NSF and from the University of Michigan. We thank M. Hayashida, Y.Y. Kovalev,
A. L\"ahteenm\"aki, \& E. Valtaoja for providing figures. The shock modeling described has made use of
data from the MOJAVE data base and from the website of the Boston U. Blazar Program.

\end{acknowledgements}

\end{document}